\documentclass[aps,onecolumn,preprintnumbers,showpacs,showkeys,nofootinbib]{revtex4-1}

\usepackage[bookmarks=false]{hyperref}
\usepackage[left=2.5cm,right=1cm,top=2cm,bottom=2cm,bindingoffset=0cm]{geometry}
\usepackage{euscript}
\usepackage{amsfonts}
\usepackage{amsmath}
\usepackage{amssymb}
\usepackage{dsfont}
\usepackage{longtable}
\usepackage{diagbox}
\usepackage{tabularx}
\usepackage{multirow}

\usepackage[dvips,pdftex]{graphicx}
\graphicspath{{figs/}}

\newcommand{\beq}{\begin{eqnarray}}
\newcommand{\eeq}{\end{eqnarray}}
\newcommand{\bew}{\begin{widetext}}
\newcommand{\eew}{\end{widetext}}
\newcommand{\be}{\begin{equation}}
\newcommand{\ee}{\end{equation}}
\newcommand{\bes}{\begin{equation*}}
\newcommand{\ees}{\end{equation*}}
\newcommand{\bdis}{\begin{displaymath}}
\newcommand{\edis}{\end{displaymath}}
\newcommand{\bga}{\begin{equation}\begin{gathered}}
\newcommand{\ega}{\end{gathered}\end{equation}}
\newcommand{\bgas}{\begin{equation*}\begin{gathered}}
\newcommand{\egas}{\end{gathered}\end{equation*}}

\begin{document}
\preprint{}

\title{Catalysis of Dynamical Chiral Symmetry Breaking by Chiral Chemical Potential in Dirac semimetals}

\author{V.~V.~Braguta}
\email[]{braguta@itep.ru}
\affiliation{Institute of Theoretical and Experimental Physics, 117259 Moscow, Russia}
\affiliation{BLTP, Joint Institute for Nuclear Research, Joliot-Curie str. 6, 141980 Dubna, Russia}
\affiliation{Moscow Institute of Physics and Technology, Institutskii per. 9, Dolgoprudny, Moscow Region, 141700 Russia}
\affiliation{Far Eastern Federal University, School of Biomedicine, 690950 Vladivostok, Russia}

\author{M.~I.~Katsnelson}
\email[]{m.katsnelson@science.ru.nl}
\affiliation{Radboud University, Institute for Molecules and Materials,
Heyendaalseweg 135, NL-6525AJ Nijmegen, The Netherlands}
\affiliation{Ural Federal University, Theoretical Physics and Applied Mathematics Department, Mira Str. 19, 620002 Ekaterinburg, Russia}

\author{A.~Yu.~Kotov}
\email[]{kotov@itep.ru}
\affiliation{Institute of Theoretical and Experimental Physics, 117259 Moscow, Russia}
\affiliation{BLTP, Joint Institute for Nuclear Research, Joliot-Curie str. 6, 141980 Dubna, Russia}
\affiliation{Moscow Institute of Physics and Technology, Institutskii per. 9, Dolgoprudny, Moscow Region, 141700 Russia}

\author{A.~M.~Trunin}
\email[]{amtrnn@gmail.com}
\affiliation{Samara National Research University, Moskovskoye shosse 34, 443086 Samara, Russia}

\begin{abstract}
In this paper we study how dynamical chiral symmetry breaking is affected by nonzero chiral chemical potential in Dirac semimetals.  To perform this study we applied lattice quantum Monte Carlo simulations of Dirac semimetals.     
Within lattice simulation we calculated the chiral condensate for various fermion masses, the chiral chemical potentials and effective coupling constants. For all parameters under consideration we have found that the chiral condensate is enhanced by chiral chemical potential. Thus our results confirms that in Dirac semimetals the chiral chemical potential plays a role of the catalyst of the dynamical chiral symmetry breaking. 
\end{abstract}

\maketitle

\section{Introduction}

The media with nonzero chiral density provide an opportunity to study a lot of interesting nontrivial physical phenomena. The most renowned example of such phenomena is chiral magnetic effect (CME)\cite{Fukushima:2008xe, Vilenkin:1980fu}, which consists in the appearance of electric current in chiral medium
along applied magnetic field. The other examples of phenomena, which take place in chiral media,
include chiral vortical effect\cite{Son:2009tf, Vilenkin:1979ui, Landsteiner:2011cp},
chiral separation effect \cite{Son:2004tq, Metlitski:2005pr}, various chiral waves\cite{Kharzeev:2010gd, Chernodub:2015gxa}. Chiral media can be created in heavy-ion collisions \cite{Kharzeev:2007jp}, in Early Universe \cite{Vilenkin:1982pn}, 
in neutron stars and supernovae \cite{Charbonneau:2009ax, Ohnishi:2014uea}. 

The chiral catalysis is one more example of the phenomenon which can be observed in media with nonzero chiral density. The mechanism responsible for this phenomenon was first explained in~\cite{Braguta:2016aov} for the theory of strong interactions~-- quantum chromodynamics (QCD).  The essence of the chiral catalysis is that nonzero chiral density generates additional fermionic states which take part in the formation of the chiral condensate. For this reason nonzero chiral density either creates or enhances the dynamical chiral symmetry breaking depending on the strength of interactions between constituents in this media. In order to study QCD with nonzero chiral density one introduces nonzero chiral chemical potential.
The influence of nonzero chiral chemical potential on the chiral symmetry breaking in QCD was considered in a number of theoretical papers~\cite{Gatto:2011wc, Chernodub:2011fr, Andrianov:2012dj, Andrianov:2013dta, Yu:2015hym, Braguta:2016aov, Khunjua:2017mkc, Khunjua:2018sro} as well as in quantum Monte-Carlo studies~\cite{Braguta:2015owi, Braguta:2015zta, Astrakhantsev:2019wnp}. Although 
all these paper consider QCD, the mechanism responsible for
the phenomenon of chiral catalysis is universal and relevant for any system with nonzero chiral density. 

Relativistic quantum field theory phenomena in condensed matter physics were intensively studied in a context 
of superfluidity of helium-3 \cite{Volovik:639409}, graphene 
\cite{27d76034ea1a423b812fdbbc9a14429a, katsnelson2012graphene} and topological insulators \cite{Qi:2011zya, Witten:2015aba}. Recent discovery of Dirac
\cite{Liu864,Neupane2014,PhysRevLett.113.027603, Liu2014AST} and Weyl \cite{Xu613,Xue1501092} Semimetals establishes additional bridges between these two fields. In particular, one can study relativistic quantum field theory with nonzero chirality. In Dirac and Weyl Semimetals nonzero chiral density and nonzero chiral chemical potential  can be created  due to the axial anomaly in parallel electric and magnetic fields\cite{Li:2014bha, Huang:2015eia}. For instance, in the Dirac semimetals Cd$_3$Ar$_2$ and Na$_3$Bi  one can generate the chiral chemical potential as large as $\sim 10-30$~meV\cite{Behrends:2015via}. 

This paper is aimed at the study of the chiral catalysis in Dirac semimetals. 
 Since we are going to consider dynamical chiral symmetry breaking which is particularly nonperturbative phenomenon, it is necessary to apply some nonperturbative approach. In our study we are going to use lattice Quantum Monte Carlo simulation, which fully accounts many-body effects in strongly coupled systems. In condensed matter physics this approach was applied in papers\cite{Drut:2008rg, Hands:2008id,Armour:2009vj, Drut:2009aj, Ulybyshev:2013swa, Boyda:2016emg, Astrakhantsev:2017isk, Boyda:2013rra, DeTar:2016vhr, Yamamoto:2016rfr, Yamamoto:2016zpx}. We have already used this approach
to study the phase diagram of Dirac Semimetals\cite{Braguta:2016vhm, Braguta:2017voo} as well as the CME in Diral Semimetals\cite{Boyda:2017dml}.  In this paper we mostly follow \cite{Braguta:2017voo} where one can find the details of the simulation. 

This paper is organized as follows. Next section is devoted to the mean field study of the dynamical chiral symmetry breaking 
in Dirac semimetal with nonzero chiral chemical potential using Nambu-Jona-Lasinio model. In Sec.~III we describe the details of the lattice simulations. In Sec.~IV we present the results of the calculation of the chiral condensate. Finally in last section we discuss our results. 

\section{Dynamical chiral symmetry breaking at nonzero chiral density}

To understand how nonzero chiral density influences dynamical chiral symmetry breaking 
we are going to use Nambu-Jona-Lasinio(NJL) model\cite{Nambu:1961tp, Nambu:1961fr} with
the Euclidean action of the form
\bew
\be
\begin{split}
S_E=\int d^4 x \biggl ( \bar \psi \bigl ( \hbar c \cdot\hat\partial   - \mu_5 \gamma_4 \gamma_5 ) \psi - G \bigr [ (\bar \psi \psi)^2 + (\bar \psi i \gamma_5 \psi)^2 \bigl ] 
\biggr ) \\ = \int d^4 x \biggl ( 
\bar \psi_R \bigl (\hbar c \cdot \hat \partial   - \mu_5 \gamma_4 ) \psi_R +
\bar \psi_L \bigl (\hbar c \cdot \hat \partial   + \mu_5 \gamma_4 ) \psi_L  
 -4 G  (\bar \psi_L \psi_R) (\bar \psi_R \psi_L)
\biggr ),
\label{eq:njl}
\end{split}
\ee
\eew
where the $\bar \psi, \psi$ are Dirac fermion fields and the $\psi_{R,L}$ are 
fermion fields with right and left chirality: $\psi_{R,L} = \bigl ( \frac {  1 \pm \gamma_5 } 2 \bigr )  \psi$, $\hat \partial=\gamma_4\frac1c\frac{\partial}{\partial t}+\gamma_i\frac{\partial}{\partial x_i}$. \footnote{In this paper we study Dirac semimetals in thermodynamic equilibrium. So, instead of real time one has Euclidean time which is designated as a fourth component of four-vector. In particular, we use the following notation $\gamma_4=\gamma_0$} The interaction between fermions in NJL model is given by the four-fermion local operator with the strength parameterized by the constant $G$. The structure of the interaction term is fixed by the requirement that action (\ref{eq:njl}) has $U_R(1)\times U_L(1)$ global chiral symmetry. 

In order to study our system at nonzero chiral density: $\langle \bar \psi \gamma_4 \gamma_5 \psi \rangle = \langle \bar \psi_R \gamma_4 \psi_R \rangle - \langle \bar \psi_L \gamma_4  \psi_L \rangle  \neq 0$, we introduced the chiral chemical potential $\mu_5$. From the second line of equation (\ref{eq:njl}) it is seen that the $\mu_5$ acts as 
 usual chemical potential $\mu=+\mu_5$ for right fermions and $\mu=-\mu_5$ for left fermions.  For this reason one can expect that in thermodynamic equilibrium nonzero $\mu_5$ leads to nonzero chiral density in the system. 

It should be noted that at sufficiently small strength of the interaction and zero
chiral chemical potential the fermion excitations of the action (\ref{eq:njl}) are Dirac fermions with the dispersion relation $E({\vec p}) =c |\vec p|$. Thus action~(\ref{eq:njl}) can be considered as low energy effective action of Dirac semimetal with one Fermi point. If the interaction is sufficiently strong and $\mu_5=0$ the system under consideration undergoes
the phase transition which dynamically breaks the chiral symmetry of the model (\ref{eq:njl}) $U_R(1)\times U_L(1) \to U_V(1)$ and forms the chiral condensate $\langle \bar \psi \psi \rangle \neq 0$ (see below). The aim of this section is to consider the model (\ref{eq:njl}) for nonzero chiral chemical potential. 

To study the phase transition in model (\ref{eq:njl}) we are going to use variational approach. It is clear that in the NJL model without interaction $G=0$, at $T=0$ and $\mu_5>0$ the vacuum state -- $| p_F \rangle$
consists of two Fermi spheres for right fermions and right antifermions with radius $\mu_5$. It is reasonable to assume that interacting 
NJL model favours condensation in the right fermion--right antifermion channel. Then a suitable vacuum state can be taken as
\begin{widetext}
\be
\begin{split}
| vac \rangle = \hat G_1 \hat G_2 \hat G_3| p_F \rangle, \qquad 
\hat G_1 =& \prod_p \biggl ( \cos {(\theta_L )} - \sin {(\theta_L )} {\hat a}^+_{L,p} {\hat b}^+_{L,-p} \biggr ), \\
\hat G_2 = \prod_{c p>\mu_5} \biggl ( \cos {(\theta_R )} + \sin {(\theta_R )} {\hat a}^+_{R,p} {\hat b}^+_{R,-p} \biggr ), \qquad
\hat G_3 =& \prod_{c p<\mu_5} \biggl ( \cos {(\tilde \theta_R )} + \sin {(\tilde \theta_R )} {\hat b}_{R,-p} {\hat a}_{R,p} \biggr ),
\end{split}
\label{vacuum}
\ee
\end{widetext}
where $({\hat a}^+_{L,p},{\hat a}_{L,p})$/ $({\hat b}^+_{L,p},{\hat b}_{L,p})$ are creation, 
annihilation operators for left fermions and antifermions, $({\hat a}^+_{R,p},{\hat a}_{R,p})$/ $({\hat b}^+_{R,p},{\hat b}_{R,p})$ are creation, 
annihilation operators for right fermions and antifermions correspondingly\footnote{Note that in the variation approach one should introduce relative phases between $\cos{(\theta)}, \sin{(\theta)}$ terms in addition to the parameters $\theta_L, \theta_R, \tilde \theta_R$. We have checked 
that trial vacuum state (\ref{vacuum}) gives minimum energy with respect to the variation over these additional phases.}. 
Here the operator $\hat G_1$ creates left states, the operators $\hat G_2$, $\hat G_3$ create right states above and below the Fermi surface correspondingly. The energy density of the state (\ref{vacuum})
can written as 
\begin{widetext}
\be
\begin{split}
E_{vac} =& 2 \int_{cp<{\mu_5}} \frac {d^3 p} {(2 \pi \hbar)^3} (cp-\mu_5) \cos^2 {\tilde \theta_R} +
2 \int_{cp>{\mu_5}} \frac {d^3 p} {(2 \pi\hbar)^3} (cp-\mu_5) \sin^2 {\theta_R} +
2 \int \frac {d^3 p} {(2 \pi \hbar)^3} (cp+\mu_5) \sin^2 {\theta_L}  \\
-& G  \biggr (  \int_{cp<{\mu_5}} \frac {d^3 p} {(2 \pi \hbar)^3} \sin {2 \tilde \theta_R} + \int_{cp>{\mu_5}} \frac {d^3 p} {(2 \pi \hbar)^3} \sin {2  \theta_R} +
\int \frac {d^3 p} {(2 \pi \hbar)^3} \sin {2 \theta_L} \biggl )^2
\end{split}
\ee
\end{widetext}
 Varying the $E_{vac}$ with respect to the parameters $\theta_L, \theta_R, \tilde \theta_R$ one obtains the following equation:
\begin{widetext}
\beq
\label{theta}
\tan {2 \theta_L} = 2 G  \frac {\Delta} {cp+\mu_5},~~~ \tan {2 \theta_R} = 2 G  \frac {\Delta} {cp-\mu_5},~~~ \tan {2\tilde \theta_R} = 2 G  \frac {\Delta} {\mu_5-cp}, \\
\Delta=\langle \bar \psi \psi \rangle= \biggr (  \int_{cp<{\mu_5}} \frac {d^3 p} {(2 \pi \hbar)^3} \sin {2 \tilde \theta_R} + \int_{cp>{\mu_5}} \frac {d^3 p} {(2 \pi \hbar)^3} \sin {2  \theta_R} +
\int \frac {d^3 p} {(2 \pi \hbar)^3} \sin {2 \theta_L} \biggl ).
\label{Mmass}
\eeq
\end{widetext}

Substituting the values of the $\theta_L, \theta_R, \tilde \theta_R$ from (\ref{theta}) to expression (\ref{Mmass}) 
one finds the gap equation with dynamical fermion mass $M=2 G \Delta/c^2$:
\bew
\beq
 \frac {1} {G} = \frac 1 {\pi^2 \hbar^3} \int_0^{\Lambda} p^2 dp \biggr [ \frac 1 {\sqrt{(c |\vec p|-\mu_5)^2+M^2 c^4}} + \frac 1 {\sqrt{(c |\vec p|+\mu_5)^2+M^2 c^4}} \biggl ]
\label{sigma1}
\eeq
\eew

Notice that the integral in the gap equation is ultraviolet divergent. Usually this integral is regularized by the three-momentum cutoff $\Lambda$ what has been done in Eq.~(\ref{sigma1}). Assuming that $Mc^2, \mu_5 \ll c\Lambda$ Eq.~(\ref{sigma1}) can be written as 
\be
\begin{split}
\frac {1} {\alpha_{NJL}} - 1 = \biggl ( y^2- \frac {x^2} 2 \biggr ) \log {\frac 1 {x^2}} 
\label{eq:gap}
\\ 
\alpha_{NJL} = \frac {G \Lambda^2} {\pi^2 \hbar^3 c },~~x =\frac {Mc} {\Lambda},~~y =\frac {\mu_5} {c\Lambda}
\end{split}
\ee

Now let us consider the system under study at zero chiral chemical potential. In this approximation Eq.~(\ref{eq:gap}) coincides with the NJL gap equation 
for dynamical fermion mass (see, for instance, review 
\cite{Klevansky:1992qe}). For  $\alpha_{NJL}<1$ left hand side of equation (\ref{eq:gap}) is positive, 
but right hand side is negative. So, there is no solution of the gap equation (\ref{eq:gap}), i.e. the $M=0$. In this case the fermion excitations have dispersion relation $E \sim |\vec p|$, the chiral condensate $\langle \bar \psi \psi \rangle =\Delta = 0$ and the system is in the semimetal phase. 

For  $\alpha_{NJL}>1$ equation (\ref{eq:gap}) has the solution $M\neq0$. In this case the fermion excitations have dispersion relation $E =\sqrt {(c {\vec p})^2 + M^2 c^4}$, there is nonzero chiral condensate $\langle \bar \psi \psi \rangle = \Delta \neq 0$ and the system is in the insulator phase. Notice that the semimetal/insulator phase transition with the described pattern was observed within lattice Monte-Carlo simulation
of Dirac semimetals in papers \cite{Braguta:2016vhm, Braguta:2017voo}.

Further let us consider how nonzero $\mu_5$ changes the properties of the gap equation. To do this we are going to study two limiting cases: the system with weakly and strongly interacting fermions. 

{\it Weakly interacting fermions.} First we are going to consider
the weakly interacting fermions: $\alpha_{NJL} \ll 1$. For $\mu_5=0$ gap equation  (\ref{eq:gap}) has no nontrivial solutions, i.e. dynamical mass is zero $M=0$. However, for any $\mu_5 \neq 0$ and $\alpha_{NJL} \ll 1$ 
there is a solution:
\beq
M^2 = (\Lambda/c)^2 \exp {\biggl [ - \frac {\pi^2 \hbar^3 c^3} {G \mu_5^2} \biggr ]}.
\label{mdyn}
\eeq
It means that even for vanishing attraction between fermions nonzero chiral chemical potential leads to dynamical chiral symmetry breaking and generation of the fermion mass.
If one rewrites the  gap equation (\ref{sigma1}), it becomes clear that the dynamical mass $M$ is determined by the behavior of the system near the Fermi surface $c|\vec k|=\mu_5$:
\begin{widetext}
\beq
 \frac {1} {G} \approx \frac 1 {\pi^2 \hbar^3} \int_0^{\Lambda} p^2 dp  \frac 1 {\sqrt{(c| p|-\mu_5)^2+M^2 c^4}}  \approx
\frac {\mu_5^2} {\pi^2 \hbar^3 c^3} \int_{\mu_5/c-\delta}^{\mu_5/c+\delta}  dp  \frac 1 {\sqrt{(|p|-\mu_5/c)^2+M^2c^2}} \approx \nu(E_F) \log {(M^2)}.
\label{sigma2}
\eeq
\end{widetext}
In the third equality we carry out the integration in the vicinity of the Fermi surface $p \in (\mu_5/c - \delta, \mu_5/c + \delta),~~\delta \ll \Lambda$. The $\nu({E_F})$ in equation (\ref{sigma2}) is a density of states on the Fermi surface
\beq
\nu({E_F}) = \frac 1 V \frac {d N(E)} {d E} \biggr |_{c |\vec p|=\mu_5} = \frac {\mu_5^2} {\pi^2 \hbar^3 c^3}.
\label{nu}
\eeq
Expression (\ref{sigma2}) represents the instability of the Bardeen-Cooper-Schrieffer theory (BCS) on the Fermi surface. Dynamical mass (\ref{mdyn}) in terms of the $\nu(E_F)$ is $M^2 = (\Lambda/c)^2 \exp { ( - 1/  {G \nu(E_F)} )}$. This expression is very similar 
to the mass gap in the BCS theory of superconductivity $\Delta = \omega_D \exp { ( - const/  {G_S  \nu_F} )}$,
where $\omega_D$ is the Debye frequency, $G_S$ is a coupling constant and $\nu_F$ is the density of states on the Fermi
surface. 

{\it Strongly interacting fermions.} Now let us proceed to the study of the gap equation (\ref{eq:gap})
in the case of strong interaction $\alpha_{NJL} \sim 1$. First we consider the case when the interaction is strong 
but insufficient for dynamical chiral symmetry breaking without the chiral chemical potential:~ $0<1-\alpha_{NJL} \ll 1$. 
As in weakly interacting medium $\mu_5 \neq 0$ leads to chiral symmetry breaking and 
generation of fermion mass:
\beq
M^2 \simeq 2 \frac {\mu_5^2} {c^4} 
\label{mass1}
\eeq

Now let us consider the case $\alpha_{NJL}>1$. In this case there is 
a solution of the gap equation (\ref{eq:gap}) for zero chiral chemical potential which we designate as $M_0$. 
For small chiral chemical potential $\mu_5 \ll M_0 c^2$ one can expand equation (\ref{eq:gap}) in 
the vicinity of the solution $M_0$ and get:
\beq
M^2 \simeq M_0^2  + 2 \frac {\mu_5^2} {c^4}.
\label{mass2}
\eeq
In the case of large chemical potential $\mu_5 \gg M_0$ dynamical fermion mass is given by formula (\ref{mass1}).

It should be noted that the term which represents logarithmic singularity
(the term $\sim y^2$ in equation (\ref{eq:gap})), i.e. dynamics near the Fermi surface,  plays a crucial role in the derivation of formulas (\ref{mass1}) and (\ref{mass2}).

The consideration conducted in this section shows that the energy minimum of the NJL model with $\mu_5 > 0$ is realized through 
the condensation of the Cooper pairs which consist of right particle and right antiparticle and break chiral symmetry. 
Notice also that in vacuum state (\ref{vacuum}) the chiral condensate is nonzero $\langle \bar \psi \psi \rangle = \Delta  \neq 0$.
The energy in the minimum is smaller than the energy of Dirac semimetal phase with free fermions. 
Thus semimetal phase is unstable with respect to chiral symmetry breaking and condensation of the Cooper pairs.

We can also conclude that the chiral chemical potential dynamically breaks chiral symmetry, if it was not broken, or strengthens it otherwise. Physically this effect stems from the formation of the Fermi surface and appearance of additional fermion states on
this surface which take part in dynamical chiral symmetry breaking. For this reason the chiral chemical potential plays 
a role of the catalyst of dynamical chiral symmetry breaking.

In this section we used NJL model which is similar to Dirac semimetal with one Fermi point and local interaction between fermions.
It is clear that this model differs from real Dirac semimetals where one has different number of Fermi points and nonlocal Coulomb interaction between fermion excitations. In this respect in real Dirac semimetals the formulas for the gap equation and the mass gap are different to that obtained in this paper. However, it is clear that in any model nonzero $\mu_5$  
leads to Fermi surface with additional fermion states which due to BCS instability create or enhance chiral symmetry breaking. 
We believe that this effect is universal and independent on number of Fermi points and interaction potential between fermions. Below we are going to study the chiral catalysis phenomenon in more realistic model applying lattice Monte Carlo simulation of Dirac semimetals.

\section{Low energy effective action for Dirac semimetals in continuum and on the lattice}

In this paper we are going to consider Dirac semimetals with two Fermi points and small isotropic Fermi velocity $v_F \ll c$. 
We believe that in its properties this theory is close to the observed Dirac semimetals Na$_3$Bi\cite{Liu864}, Cd$_3$As$_2$\cite{Neupane2014,PhysRevLett.113.027603}. 
Low energy effective theory of fermionic excitations in Dirac semimetals can be described by two flavors of 3D Dirac fermions.
Due to the smallness of the Fermi velocity magnetic interactions and retardation effects can be safely disregarded. As the result the interaction in Dirac semimetals is reduced to instantaneous
Coulomb potential. Taking into account these properties it is straightforward to build the partition function for Dirac semimetals:
\beq
Z=\int D\psi~ D\bar\psi~DA_4~\exp{\bigl ( -S_E \bigr )},
\label{eq:Z}
\eeq
where $\bar \psi, \psi$ are Dirac fermion fields, $A_4$ is temporal component of the vector potential of the electromagnetic field.
The Euclidean action $S_E$ can be written as a sum of the contributions of the gauge $S_{g}$ and the fermion fields $S_f$ 
\beq
S_E=S_f+S_g.
\label{eq:continuousaction}
\eeq
The gauge and fermionic parts of the action can be written in the following form
\beq
S_g &=& \frac{1}{2 e^2}\int d^3x dt (\partial_i A_4)^2, 
\nonumber \\
S_f &=& \sum\limits_{a=1}^{N_f}\int d^3x dt\, \bar{\psi}_a \left(\gamma_4(\hbar \partial_4+iA_4)+ \hbar v_{F}\gamma_i\partial_i+\right.\nonumber\\
&&\qquad\qquad\qquad\qquad\qquad+\left.\mu_5\gamma_5\gamma_4\right)\psi_a 
\label{eq:sf}
\\  
&=& \sum\limits_{a=1}^{N_f}\int d^3x dt\, \bar{\psi}_a D_a(A_4) \psi_a\nonumber .
\eeq
The field $A_4$ in equations (\ref{eq:sf}) can be integrated out thus leading to the theory where Dirac fermions interact via the instantaneous Coulomb law with the effective coupling constant $\alpha_{eff}=e^2/4\pi \hbar v_F$.

Note that in the fermion part of the action we have introduced the chiral chemical potential $\mu_5$, which leads to nonzero chiral density in the system under investigation.  We have already mentioned that nonzero chiral chemical potential 
can be created due to the axial anomaly in parallel electric $\vec E$ and magnetic $\vec B$ fields. At zero temperature the value of the chiral chemical potential can be estimated using formulas from paper \cite{Li:2014bha} 
\beq
\mu_5 = \hbar v_F \biggl ( \frac 3 4  \frac {e^2} {\hbar^2 c}  {\vec E \cdot \vec B} \tau \biggr )^{1/3},
\label{CCP}
\eeq
where $\tau$ is the relaxation time
of chiral charge.
In the Dirac semimetals Cd$_3$Ar$_2$ and Na$_3$Bi  one can generate the chiral chemical potential as large as $\sim 10-30$~meV\cite{Behrends:2015via}.

Further one can perform the standard rescaling of the time coordinate and timelike component of the electromagnetic field: $t\to t/ v_{F}$, $A_4\to v_F A_4$. It leads to the following modification of the action:
\beq
S_g &=& \frac{v_F}{2e^2}\int d^3x dt (\partial_i A_4)^2, \nonumber \\
S_f &=& \sum\limits_{a=1}^{N_f}\int d^3x dt\, \bar{\psi}_a \left(\gamma_4(\hbar \partial_4+iA_4)+ \gamma_i\cdot \hbar \partial_i+\right.\nonumber\\
&&\qquad\qquad\qquad\qquad\qquad+\left.\frac{\mu_5}{v_F}\gamma_5\gamma_4\right)\psi_a,
\eeq

For the discovered Dirac semimetals $v_F\ll c$ and $\alpha_{eff}>1$. Thus the system is strongly coupled and one should apply nonperturbative methods for studying such systems. In this paper we are going to apply lattice Quantum Monte Carlo simulations. For numerical study we employ the staggered discretization for fermions and noncompact discretization of lattice gauge fields~\cite{Braguta:2017voo}. 

To write a discretized version of action (\ref{eq:Z}) 
we introduce a regular cubic lattice in four dimensional space with spatial lattice spacing 
$a_s$ and temporal lattice spacing $a_t$ ($a_t=\xi a_s$). As discussed in \cite{Braguta:2017voo}, it is important to take the limit $\xi=a_t/a_s\to0$. The number of lattice sites is $L_s$ in each spatial direction
and $L_t$ in temporal direction. 
For lattice simulation we will use Lorentz-Heaviside units and it will be assumed that $\hbar=c=1$. In addition we will take $a_s=1$, restoring explicit spatial lattice spacing when necessary.
The resulting action can be written as
\bes
S_G=\frac{1}{8\pi\alpha_\text{eff}}\sum_{x}\sum_{i=1}^3(\theta_{4,x}-\theta_{4,x+\hat\imath})^2,
\ees
\be
\label{eq:actionf}
\begin{split}
S_F=&\sum_x \left( m \xi \,\bar\psi_x\psi_x
+\frac{\xi}{2}\sum_{i=1}^3\eta_i(x)\bigl[\bar\psi_x\psi_{x+\hat\imath}-\bar\psi_{x+\hat\imath}\psi_{x}\bigr]
\right.\\
&+\left.\frac{1}2\eta_4(x)\bigl[\bar\psi_xe^{i\theta_4,x}\psi_{x+\hat4}-\bar\psi_{x+\hat4}e^{-i\theta_4,x}\psi_{x}\bigr]\right.\\
&+\left.\frac12\, \frac {\mu_5 a_t} {v_F}
\sum_xs(x)\bigl[\bar\psi_{x+\delta}\psi_{x}-\bar\psi_x\psi_{x+\delta}\bigr]\right),
\end{split}
\ee
where gauge~$\theta_{4,x}$ and fermionic~$\psi_x$ fields are defined at the sites $x=(x_1,x_2,x_3,x_4)$ of Euclidean lattice. The factors $\eta_1(x)=1$ and $\eta_{\mu>1}(x)=(-1)^{x_1+\ldots +x_{\mu-1}}$ are the standard staggered phase factor. Similarly to \cite{Braguta:2015zta} the chiral chemical potential~${{\tilde\mu}_5}$ is introduced through the additive term in the last line of Eq.~\eqref{eq:actionf}, where $s(x)=(-1)^{x_2}$ and $x+\delta=\{x_1+1,x_2+1,x_3+1,x_4\}$. 
Notice that the lattice parameter $m$ from lattice action (\ref{eq:actionf}) is related to it physical values as $m=m_{phys} a_s/v_F$.
Notice also that below we are going to write the chiral chemical potential in lattice units ${\tilde\mu}_5 = \mu_5 a_s/v_F$.

It should be noted here that the direct lattice simulation at $m=0$ is impossible due to the irreversibility of the Dirac operator and related numerical instabilities.
For this reason the simulations are performed at several small but non-zero values of $m$ and observables are subsequently extrapolated to the chiral limit(see below).

Diagonalizing lattice action~\eqref{eq:actionf} it can be shown that it corresponds to $N_f=4$ flavors of Dirac fermions\cite{Braguta:2015zta}.
To obtain $N_f=2$ flavors of Dirac fermions we take a square root of the fermion determinant. 

It is interesting to estimate the lattice spacing $a_s$. It is clear that inverse lattice spacing is of the order of ultraviolet cut off in the effective theory (\ref{eq:sf}), i.e. 
of the order of the inverse characteristic distance between atoms in Dirac semimetal under consideration\footnote{Notice that for graphene this was explicitly demonstrated in paper \cite{Astrakhantsev:2015cla}}. The characteristic distance between atoms in Cd$_3$Ar$_2$\cite{Liu2014AST} and Na$_3$Bi\cite{Liu864} is $\sim 5~\AA$. So,
taking typical Fermi velocity in Dirac semimetal $v_F/c \sim 0.001$ 
one obtains the relation $\hbar v_F/a_s \sim 0.4~\mbox{eV}$  which can be used to estimate 
physical values of different dimensional parameters. In particular, characteristic value of 
the fermion mass used in the simulation is $m = m_{phys} a_s/v_F \sim 0.1$ (see, below). So, 
in physical units this mass is $m_{phys} c^2 \sim 0.04~$eV.
We have already mentioned that in paper \cite{Behrends:2015via} it was shown that one can generate the chiral chemical potential as large as $\sim 20$~meV. It causes no difficulties to calculate this value of the $\mu_5$ in lattice units: $\tilde \mu_5 \sim 0.05$.

\begin{figure*}[t!]
\begin{center}
\includegraphics{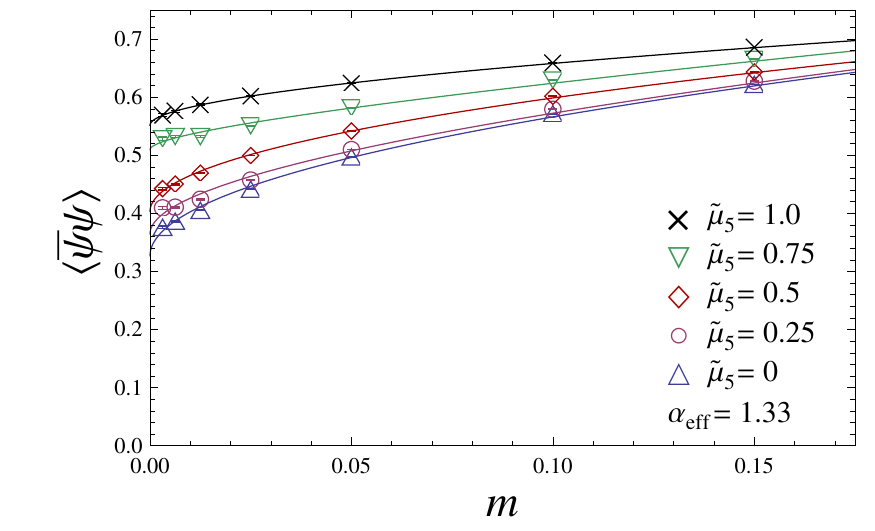}\includegraphics{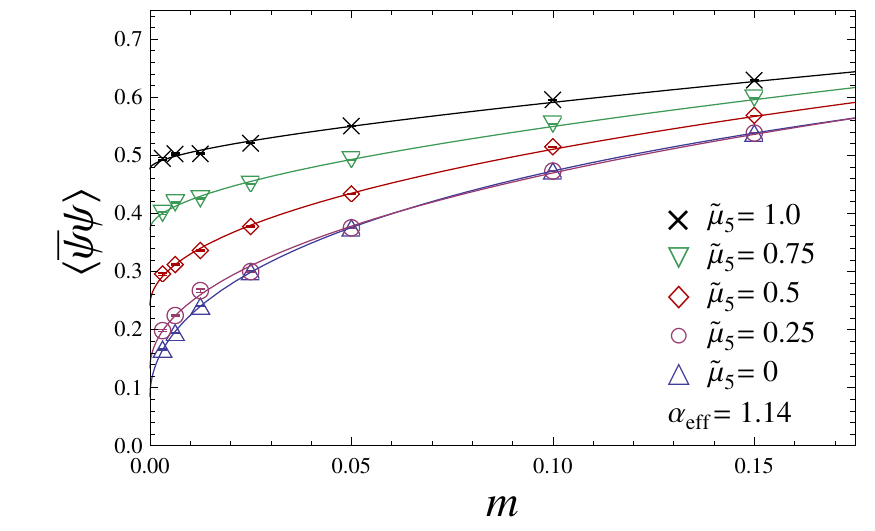}
\includegraphics{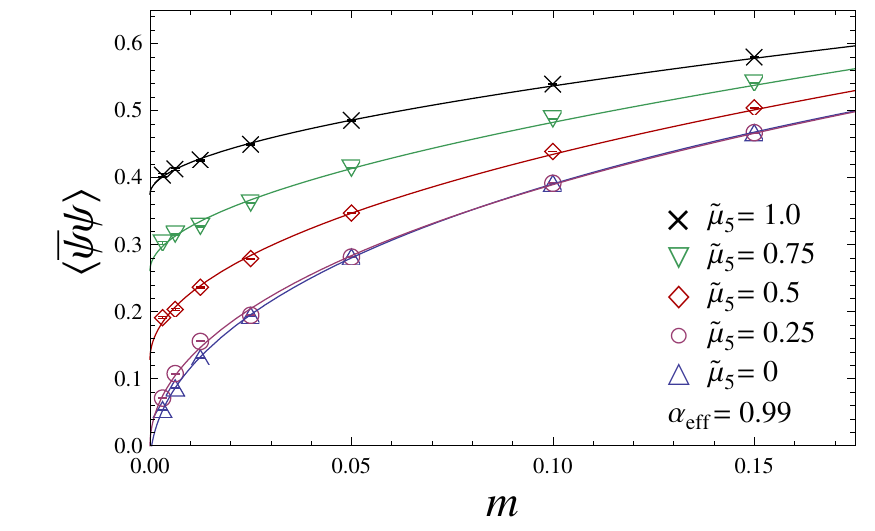}\includegraphics{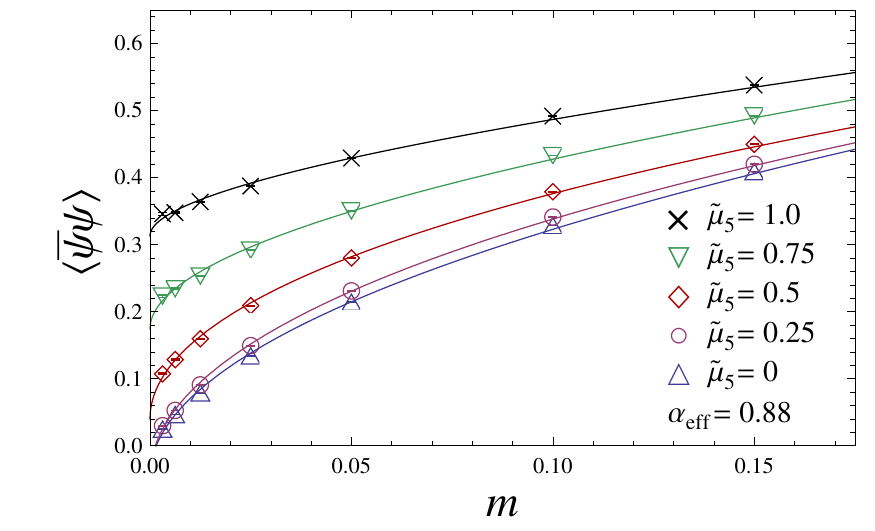}
\end{center}
\caption{The chiral condensate $\langle \bar \psi \psi \rangle$ as a function of the fermion mass  for various ${\tilde\mu}_5$ and $\alpha_{eff}$. The fermion mass and the ${\tilde\mu}_5$ are shown in lattice units.  The lattice size is $80\!\times\!16^3$ and the anisotropy is $\xi=1/5$.  The mass dependence of the chiral condensate was fitted by Eq.~\eqref{eq:chi-fit}.
$\alpha_\text{eff}=1.33$ ({\itshape top left}) is in the insulator phase, $\alpha_\text{eff}=1.14$ ({\itshape top right}) is near the semimetal-insulator phase transition, $\alpha_\text{eff}=0.99$ ({\itshape bottom left}) and $\alpha_\text{eff}=0.8$ ({\itshape bottom right}) correspond to the semimetal phase at ${\tilde\mu}_5=0$.
}
\label{fig:pbp-vs-ma-30x10} 
\end{figure*}

\begin{figure}[t]
\begin{center}
\includegraphics{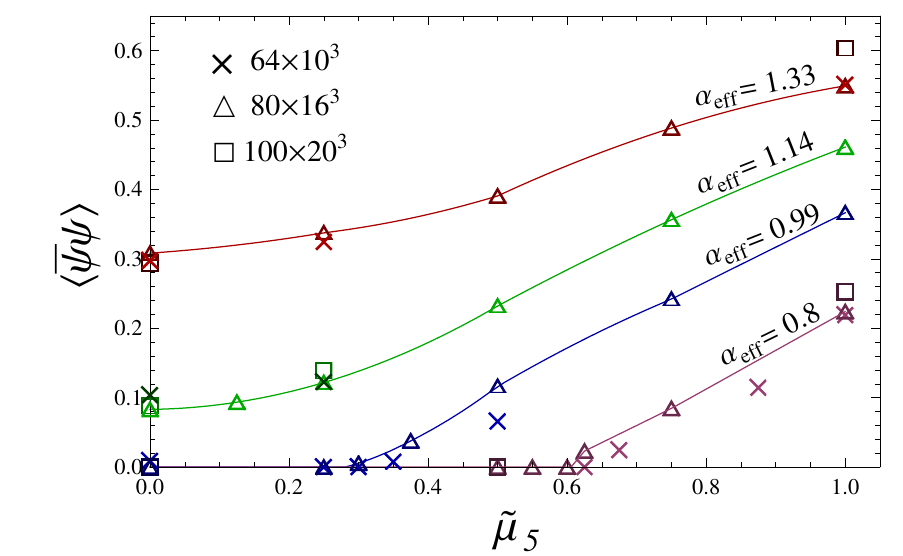}
\end{center}
\caption{The chiral condensate as a function of the ${\tilde\mu}_5$ calculated on different lattices and for various effective coupling constants. The curves are to guide the eyes.
}
\label{fig:pbp-vs-mu5} 
\end{figure}

\section{The chiral condensate}
In this section we apply lattice Monte Carlo approach in order to address the question how nonzero chiral chemical potential influences dynamical chiral symmetry breaking in the system with action (\ref{eq:Z}). The order parameter for the the chiral symmetry breaking/restoration transition is the chiral condensate which can be calculated on the lattice as  
\be
\label{eq:chi-def}
\langle \bar \psi \psi \rangle=- \frac{1}{4}\langle  S(x,x)\rangle,
\ee
where $S(x,y)$ is the fermion propagator in theory (\ref{eq:actionf}). Notice that action (\ref{eq:actionf}) after diagonalization corresponds to four Fermi points or four fermion flavours. The factor 1/4 in last equation is introduced to calculate the chiral condensate per one flavour.  

In the chiral limit, the vanishing condensate $\langle \bar \psi \psi \rangle(m\to0)=0$ corresponds to the chirally symmetric phase and the system is a semimetal, while for  $\langle \bar \psi \psi \rangle(m\to0) \neq 0$ the chiral symmetry is broken and the system becomes an insulator. In our previous study~\cite{Braguta:2017voo} with the action (\ref{eq:actionf}) we found that at zero chiral chemical potential this phase transition takes place at $\alpha_{c,\mu_5=0}\approx1.1$. Now let us consider what happens at nonzero $\mu_5$. 

We have already mentioned that it is not possible to perform lattice simulation at $m=0$  due to the irreversibility of the Dirac operator and related numerical instabilities.
So, the simulations are performed at several small but non-zero values of $m$ and observables are subsequently extrapolated to the chiral limit. In the extrapolation
we used the following ansatz
\be
\label{eq:chi-fit}
f(m)=k_0+k_1\sqrt{m}+k_2m,
\ee
which proved its applicability in QCD~\cite{Ejiri:2009ac,Bazavov:2011nk,Braguta:2015zta}.

We proceed to the results of simulation on $80\!\times\!16^3$ lattice. In paper Ref.~\cite{Braguta:2017voo} it was shown that the limit $a_t\to0$ is crucial for obtaining correct lattice theory of Dirac semimetals and reducing finite-volume artifacts. So, in practical simulations the lattice spacing in temporal direction has to be taken at least several times smaller than its spatial counterpart. In the simulation on the lattice $80\!\times\!16^3$
we fixed the anisotropy ratio $\xi=a_t/a_s=1/5$. In paper \cite{Braguta:2017voo} it was shown that this ratio is rather good approximation to the limit $a_t\to0$. 

In Fig.~\ref{fig:pbp-vs-ma-30x10} we show the chiral condensate $\langle \bar \psi \psi \rangle$ as a function of the fermion mass  for various ${\tilde\mu}_5$ and $\alpha_{eff}$. The fermion mass is shown in lattice units. For the calculation we chose four values of the effective coupling constant:
$\alpha_\text{eff}=1.33$, 1.14, 0.99, 0.8. 
At zero chiral chemical potential 
the $\alpha_\text{eff}=1.33$ correspond to the insulator phase,
the $\alpha_\text{eff}=1.14$ above but close to the semimetal/insulator phase transition and the $\alpha_\text{eff}=0.99, 0.8$ are in semimetal phase. 
The mass dependence of the chiral condensate was fitted by Eq.~\eqref{eq:chi-fit}.

It is seen that for all values of the simulation parameters the chiral chemical potential always enhances the chiral symmetry breaking in the system, what confirms the chiral catalysis phenomenon in Dirac semimetals. 

In Fig.~\ref{fig:pbp-vs-mu5} we plot the chiral condensate $\langle \bar \psi \psi \rangle$ in the chiral limit as a function of the ${\tilde\mu}_5$. From this figure one sees that 
in the insulator phase the chiral chemical potential enhances the chiral condensate what agrees with the mean-field consideration performed in Sec.~II. If the system in the semimetal phase the chiral condensate is zero up to some value of the chiral chemical potential ${\tilde\mu}_5<{\tilde\mu}_5^c$ and develops nonzero value in the region ${\tilde\mu}_5>{\tilde\mu}_5^c$ where the $\langle \bar \psi \psi \rangle$ is rising function of the $\mu_5$. The rise of the chiral condensate in the chiral limit with $\mu_5$ again confirms the chiral catalysis phenomenon in Dirac semimetals.

\begin{figure}[t]
\begin{center}
\includegraphics[scale=0.9]{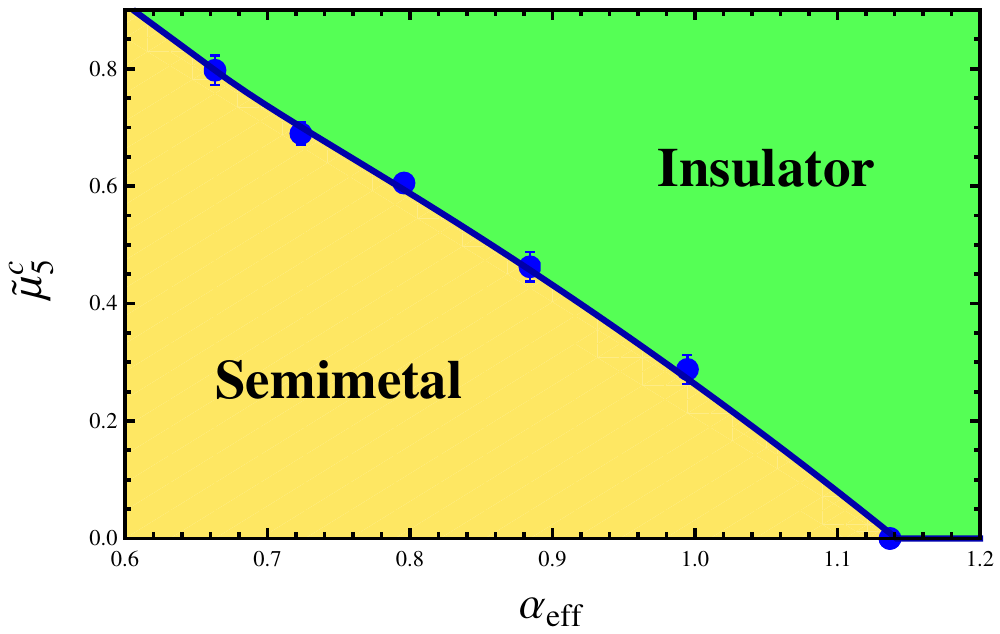}
\end{center}
\caption{The dependence of critical ${\tilde\mu}_5^c$ on effective coupling. The curve is to guide the eyes.
}
\label{fig:mu5-beta} 
\end{figure}

Our results imply that for any $\alpha_{eff}$ in the semimetal phase  there exists the critical value of the chiral chemical potential ${\tilde\mu}_5^c$ after which the system under study turns into insulator phase. We calculated the 
${\tilde\mu}_5^c$ for various $\alpha_{eff}$. In  Fig.~\ref{fig:mu5-beta} we show how the ${\tilde\mu}_5^c$ depends on the effective coupling constant. 

From Fig.~\ref{fig:mu5-beta} it is seen that the larger the chiral chemical potential the smaller critical effective coupling constant after which the system transfer from the semimetal to the insulator phase. This observation is also in agreement with the chiral catalysis phenomenon. Notice, however, that the existence of the critical ${\tilde\mu}_5^c$ seems to contradict to the mean-field study. In Sec.~II it was predicted that for any value of the $\alpha_{eff}$ nonzero chiral chemical potential leads to chiral symmetry breaking, i.e. there is no ${\tilde\mu}_5^c$. The contradiction can be resolved if we recall that we conduct our study at finite $L_t$, i.e. small but finite temperature effects are present in our analysis. At the same time the study of Sec.~II is performed at zero temperature. Notice also that the mass gap in the weak coupling regime what corresponds to the semimetal phase is exponentially suppressed (\ref{mdyn}). Evidently such small mass gap will be destroyed by finite temperature effects. In other words  to observe exponentially small mass gap one needs huge number $L_t$ what is not accessible in our present analysis. 
To summarize, finite temperature leads to appearance of the critical chiral chemical potential ${\tilde\mu}_5^c$ which depends on the effective coupling constant and where the system turns from the semimetal to the insulator phase. 

We would like also to notice that the magnetic catalysis phenomenon~\cite{Gusynin:1994re, Gusynin:1995nb} at finite temperature looks similar to the chiral catalysis. In this case 
nonzero magnetic field leads to dynamical chiral symmetry breaking for any effective coupling similarly to $\mu_5$ in the chiral catalysis. However, in lattice simulation of graphene~\cite{Boyda:2013rra} in external magnetic field the phase diagram looks similar to Fig.~\ref{fig:mu5-beta}.

At the end of this section let us study  finite-volume effects. To do this we 
performed numerical simulation on the lattices
 $64\!\times\!10^3$ and $100\!\times\!20^3$. 
 The results of these additional measurements are presented in Fig.~\ref{fig:pbp-vs-mu5}. It is seen from this figure that the results obtained at different lattices are agreement with each other i.e. the finite-volume effects do not affect our results.

\section{Conclusion}

The aim of this paper is to study the chiral catalysis phenomenon in Dirac semimetals with nonzero chiral density. The essence of the chiral catalysis is that nonzero chiral density generates additional fermionic states which take part in the formation of the chiral condensate. For this reason nonzero chiral density either creates or enhances the dynamical chiral symmetry breaking depending on the strength of interactions between constituents in this media. In order to create nonzero chiral density in the system under study we introduce nonzero chiral chemical potential. To perform this study we applied lattice quantum Monte Carlo simulations of Dirac semimetals, which fully accounts many-body effects in
strongly coupled systems.   

We calculated the chiral condensate for various fermion masses, chiral chemical potentials and effective coupling constants. For all parameters under study we have found that the chiral condensate rises with rising chiral chemical potential.  This confirms that in Dirac semimetals the chiral chemical potential plays a role of the catalyst of the dynamical chiral symmetry breaking. 

We also calculated finite temperature phase diagram for the Dirac semimetals in the plane effective coupling constant--chiral chemical potential. We have found that the larger the chiral chemical potential the smaller critical effective coupling constant after which the system turns from the semimetal to the insulator phase. This observation is also in agreement with the chiral catalysis phenomenon. 

In paper \cite{Behrends:2015via} it was shown that one can generate the chiral chemical potential as large as $\sim 10-30$~meV for magnetic field $|\vec B|=1~$mT. In lattice units this corresponds to the $\tilde \mu_5 \sim 0.03-0.08$. From Fig.~\ref{fig:pbp-vs-ma-30x10} it is seen that the effect to the system of such chiral chemical potential is quite small. Notice, however, that we have carried out rather rough estimation of the lattice spacing, i.e. the chiral chemical potential in lattice units. In real Dirac Semimetals the chiral chemical potential in lattice units might be larger and the effect might be more pronounced. Notice also that according to formula (\ref{CCP}) to generate larger chiral chemical potential one can 
use larger magnetic field or larger relaxation time of chiral charge. 
Using high magnetic fields ($> 10$~T) one can increase the chiral 
chemical potential in more than an order of magnitude which seems to be 
enough to reach the catalysis regime. The other option is an increase of 
the chiral relaxation time due to the use of very clean samples and low 
temperatures. Therefore we believe that the predicted effects can be 
experimentally observable in real materials.

\section{ACKNOWLEDGMENTS}

The work of V.\,V.\,B. and A.\,Yu.\,K., which consisted of 
processing and physical interpretation of lattice data, was
supported by grant from the Russian Science Foundation
(project number 16-12-10059). 
A.\,M.\,T. acknowledge the support from the RFBR grant with the number 18-02-01107. This work has been carried out using computing resources of the federal collective usage center Complex for Simulation and Data Processing for Mega-science Facilities at NRC ``Kurchatov Institute'',~\url{http://ckp.nrcki.ru/}. In addition, the authors used the equipment of the cluster of the Institute for Theoretical and Experimental Physics and the supercomputer of Joint Institute for Nuclear Research ``Govorun''.

\end{document}